# Leveling the playing field for Visually Impaired using Transport Assistant

**Gourav G. Shenoy, Mangirish A. Wagle, Kay Connelly**
Indiana University School of Informatics and Computing
Bloomington, IN 47405 USA
{goshenoy, mawagle, connelly}@indiana.edu

**ABSTRACT**
Visually impaired people face numerous challenges when it comes to transportation. Not only must they circumvent obstacles while navigating, but they also need access to essential information related to available public transport, up-to-date weather forecast, and convenient method for booking private taxis. In this paper we introduce *Transport Assistant* - a voice based assistive technology prototype, built with a goal of leveling the playing field for the visually impaired to solve these problems that they face in their day to day life. Being voice enabled makes it seamlessly integrate into the environment, and can be invoked by saying a hotword "hello assistant". The paper explores this research question, followed by investigating existing technologies, explains the methodology and design, then concludes by presenting the prototype and results.

**ACM Classification Keywords**
H.5.2. Information interfaces and presentation: User Interfaces - Voice I/O.

## 1. INTRODUCTION

Computer technology is advancing at exponential speeds, as computing devices are becoming progressively smaller and powerful. The idea that this technology can be embedded in everyday devices, also known as Ubiquitous Computing, is often mentioned in the context of improving healthcare. Visual impairment is a term experts use to describe any kind of vision loss, whether it's someone who cannot see anything at all or someone who has partial vision loss.

Visually impaired people confront a number of visual challenges everyday[8], from reading the label on a frozen dinner to figuring out if they're at the right bus stop. We conducted research to find out more about the various issues that visually impaired people face in other domains. We started by reading some very informative literature about visual impairment and existing technologies that are in place to help some of the problems that visually impaired people face. We conducted in-person interviews with the target population to get first hand information about the daily challenges they face, how open they were to using technology, and what expectations they had from technology. We also sought expert views and advice to get more insights.

To better visualize and consider our target population, we created 2 personas from the information we collected, which pointed out the need for technology to solve problems related to privacy, mobility and vision. We used these personas as a design tool during our brainstorming. Based on the interview data, feedback and analysis, it was realized that transportation is one of the less attended problems faced by visually impaired. Hence we chose to address the key concerns with respect to transportation through this project.

Also, providing the visually impaired individuals with an advanced and effective transportation assistance tool will be significant in the following three aspects: First, it will reduce some of the suffering that people with visual disabilities face. Second, it will help these people to live independently. Third, it will promote employment, benefiting the society by fully utilizing the talents and abilities of this portion of the population.

**Research Question and Sub-problems**

Our research question is, "Is voice-based travel assistance technology a practical replacement to the cane stick or service dog used by visually impaired people to navigate?"

Our first research sub-problem began by conducting interviews with the visually impaired community. This was a critical part of our research, as it had to cater to their needs. The feedback we received from them provided healthy inputs which proved to be crucial in designing our prototype.

Our second research sub-problem focused on deciding the most intuitive way to design a "pervasive" prototype. For visually impaired population, a voice based communication system seemed the best option, which would seamlessly blend into their lives.

As part of our third research sub-problem we focused on deciding cost effective and interoperable components. Since most of the visually impaired population live in low-income settings, an economically viable solution is essential. Also, it is very important to achieve interoperability between the different components.

Finally our fourth research sub-problem dealt with using visual recognition services, such as IBM Watson Visual Recognition APIs, to perform obstacle detection and avoidance. These APIs provide ready to use machine learning classifiers, which can be extended to identify other items by explicit training, and used to alert the visually impaired when approaching a zebra crossing while navigating outdoors; or alerting when an obstacle is in the way while navigating indoors.

**Problem Setting**
According to the World Health Organization (WHO) 285 million people are estimated to be visually impaired: 39 million are blind and 246 million have low vision [9]. Moreover, ninety percent of the population live in a low income setting. Most of the visually impaired people depend on a cane stick, a well trained service dog, or a good friend to help them navigate, along with helping perform other day to day activities [10]. There are technologies out there today which are targeted towards solving discrete problems, such as object recognition, map based navigation, voice search, and so on.

There are also other options such as Google Glass [5], and augmented reality devices for performing obstacle detection. Nevertheless there are variety of obstacle avoidance mechanisms that can be used such as electromagnetic tracking devices to identify obstacles, RF (Radio Frequency) localization, or ultrasonic SONAR (SOund Navigation and Ranging) sensors [11]. None of these techniques, if used independently can offer a concrete solution to aid the visually impaired.

Through our project we aim to build an assimilated solution for transportation related problems of the visually impaired, which will be pervasive, cost effective, and provide a conversational service. An important factor in designing any new system is to make sure the user experience is good, along with making sure the results are accurate. Therefore, fast retrieval of information and enhanced user experience are also prime objectives of our prototype.

**EXISTING TECHNOLOGIES**
In this section we include some of the existing assistive technologies available for visually impaired and blind people. Some of these technologies have been well developed, such as speech to text, and visual recognition. Therefore, by investigating these technologies helped us understand how we can integrate, and not re-develop them in this research, to solve the bigger issue of transportation among the visually impaired.

**Audio Assistance and Speech Recognition Technology**
The Google Voice Search is an excellent example of an audio based system. Proper training helps the system to become familiar with the voice of the user. This training improves the accuracy of speech recognition significantly. There are many applications of the software system, such as voice search on Google maps and Google mobile apps. This system has already been integrated into Android, which is the operating system in many smartphone devices. The accuracy of the Google voice search is excellent, approaching hundred percent in a quiet environment, and the performance is satisfactory even when a certain amount of noise is present.

The text-to-voice technique is also available. One can get clear and loud navigation instructions from these devices with almost real-time performance. Some other systems like computer assistant answering machine and Google voice mailbox also provide similar functions. Software is available online to facilitate practical applications.

One ideal example of combining the voice search and text-to-voice techniques is Siri, shown in Figure 1, for iOS, the operating system of Apple's iPhone. One can talk to Siri and get a human-like response in voice. This system can be used for automatic voice assistance in daily life with both voice input and synthesized voice output.

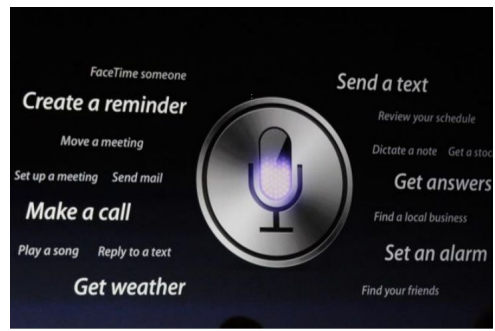

**Figure 1. Siri from Apple's iOS [11]**

**Visual Recognition Services**
The IBM Watson Visual Recognition service, analyzes the image provided and returns a list of weighted keywords, which it transforms into a sentence. These Application Programming Interfaces (APIs), together with text to speech APIs can be used to build cognitive applications for blind people. These APIs are available on IBM Watson Developer Cloud on Bluemix, which gives developers the ability to build cognitive applications of the future today.

IBM has also added tooling capabilities and enhanced Software Development Kits (SDKs) in multiple programming languages, making it easier and faster for developers to customize and build with several Watson cognitive and mobile services, as shown in Figure 2.

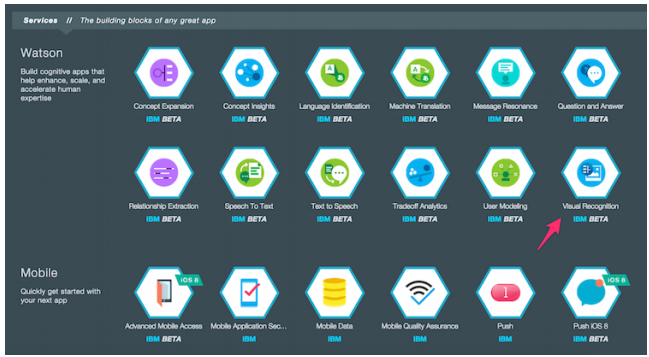

**Figure 2. List of IBM Watson cognitive services available on IBM Bluemix Platform-as-a-Service (PaaS)**

## METHODOLOGY

As part of this research project, we read a lot of relevant literature, conducted several interviews with visually impaired people here at Indiana University Bloomington, and met experts in this field to get their valuable inputs. We also regularly conducted group discussion sessions to brainstorm several design ideas and chalked them out on a whiteboard. These ideas were targeted towards solving either one or many of the problems faced by visually impaired people; keeping in mind the feedback. We neatly categorized all design ideas based on the problem domain they fall into. Some of them include:

- Mobility: assistive technology for helping with navigation and transportation related issues.
- Security: assistive technology to prevent privacy attacks such as shoulder surfing, and burglary.
- Social: technology for helping visually impaired enjoy a better social life; eg: recognizing friends, playing games, avoiding social stigma.
- Household: assistive technology to help with tasks such as finding items or medicines inside the house, and assists with cooking, etc.
- Employment: assistive technology to help with daily employment tasks such as preparing presentations, etc.

We then discussed these ideas with Prof. Connelly and sought opinions from expert Tousif Ahmed (Eshan), PhD student at School Of Informatics and Computing who has worked on significant research being done towards the physical privacy issues of visually impaired people [3,4]. Based on the feedback we shortlisted 3 ideas:

1. Transport Assistant: assistive technology that helps with all aspects related to transportation.
2. Read Out Aloud: glasses that can scan text on any surface and read it out aloud.
3. Take-Me Band: A wristband that provides feedback through vibrations to detect any obstacles and circumvent it, while navigating to requested destination.

We mapped these ideas to the requirements gathered from the target population, created storyboards, and presented it to students in the I527 class who provided healthy comments and suggestions along with some constructive criticism. Several factors motivated us to select *Transport Assistant* as the final design idea, including the fact that transportation technology was an area among visually impaired that demanded significant improvement. Figure 3, shows the storyboard for the *Transport Assistant* design idea.

We also had the privilege to meet the Director of Disability Services for Students (DSS), Shirley Stumpner and the interaction with her was quite encouraging and informative. She gave us insights to the common problems that are addressed by DSS, and pointed out to some organizations which provide newspaper reading assistance to visually impaired, like NFB Newsline. The Directory of DSS also helped us to connect with visually impaired students on campus; we are grateful to them for being available to interview with us and sharing their problems and thoughts.

Some of the challenges we identified were, how can we build a pervasive technology that would work seamlessly to cater to the transportation problems of the visually impaired, along with addressing the issue of social stigma that is usually faced by the blind and also indicated in the research papers that we read [2,6,7,8]. Also, can this idea be extended to allow users to get "any" information, not just transportation related? These parameters play a very decisive role in coming up with the right design for such an assistive application.

## DESIGN

As mentioned earlier, the feedback we received from multiple sources - interviews, expert opinions, storyboard presentations - provided us with helpful parameters, based on which we could tune the design of our final prototype. One of the most important requirements of ubiquitous [1] computing is the need to be seamlessly integrated into the user's environment, such that the user does not realize that he/she is consuming the service. The most intuitive way to achieve this for visually impaired is to have a voice-based interaction model, which would act like a conversation agent by responding to user's questions in human-like tone.

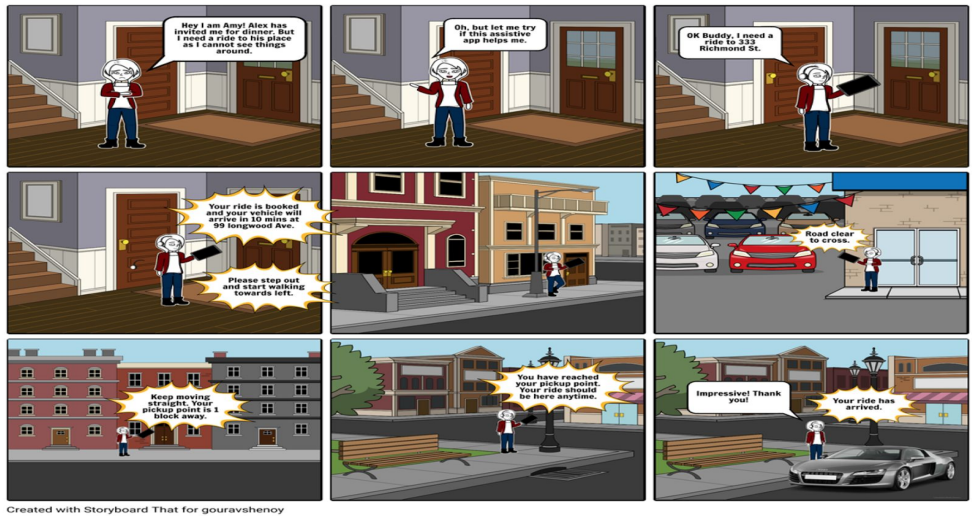

**Figure 3. Storyboard that depicts the key scenario of the application assisting a visually impaired to order a ride**

We decided that the Transport Assistant would be a service application running on an Android phone. This service needs to be "always listening" even while running in the background, and react promptly when triggered by a *hotword* - similar to "Ok Google" in Android, or "Siri" in iOS. We realize that all electronic devices have battery limitations, therefore we have made sure our application is battery efficient [12], by designing the code quality similar to Google Now. Keeping this limitation in mind we also decided to extend our design to detect low battery, and give out alerts to the user to find an alternative source or seek help.

We have broken down the design of *Transport Assistant* in multiple sub-sections to get a better understanding of the physical design (what components make up the assistant), interaction model (how these different components interact), system use (use-cases of transport assistant), and technical requirements.

**Physical Design**

There are 5 main entities that make up the physical design for *Transport Assistant*. These entities are:

1. User - a visually impaired person using the Transport Assistant.
2. Microphone - receive voice commands from the user (either for navigation, or general information queries).
3. Headphone - provide audio output to the user; this could be either real time navigation instructions, or answer to the question asked by the user.
4. Camera - asynchronously fetch visual information, which will be used to identify and circumvent obstacles during navigation. After much thought, we decided to consider mounting the camera on the waist to obtain more stable image stream, in contrast to head mounted camera which would have a lot of irregular movements.
5. Server application - processes requests from the phone client and responds with appropriate information (navigation, or general Q&As).
6. Android phone - runs the transport assistant client application which talks to the transport assistant server application.

Figure 4 is a sketch that puts together all these entities in the form of a visual representation of the physical design for the Transport Assistant. Also the camera, and headphone (with mic) are bluetooth or WiFi enabled and use the same protocol to communicate with the client running on the android phone.

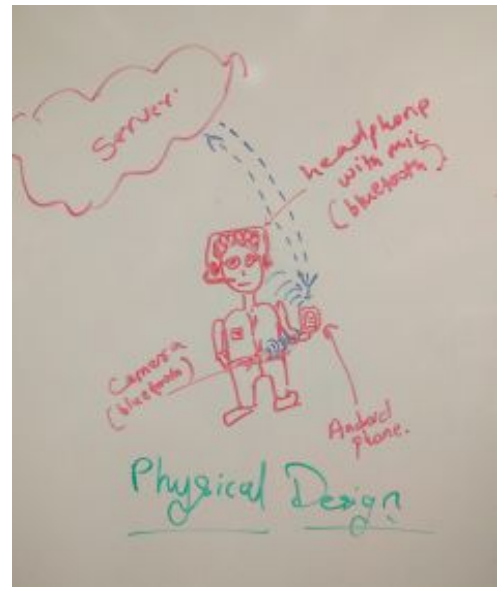

**Figure 4. Physical design sketch for Transport Assistant**

**Interaction Model**

Figure 5 depicts the sequence of interactions between the different entities of the Transport Assistant for a typical use-case.

There would be two types of flows:

User requesting for the transport (Synchronous events).

1. User issues a voice command to request a transport to destination using a microphone.
2. Microphone relays it to an Android phone which then forwards the request to the server.
3. The server processes the request and responds with the destination navigation information to the phone.
4. The phone would then provide options to user whether he / she wants to use public or private transport. The options are provided to user using voice through headphones.
5. Depending on the choice of user the phone would then relay the voice alerts and navigation instructions to the user through the headphones.

The system collecting image data and location to generate voice alerts (Asynchronous events)

1. The transport assistant client application running on the phone is programmed to use the camera to asynchronously click images.
2. The images along with the location data would be sent to the server.
3. Server would then process the image data and location to generate navigation instructions and alerts on the fly, which it would send back to the phone client application as response.
4. The phone would then narrate the instructions and alerts to the user using voice through headphones.

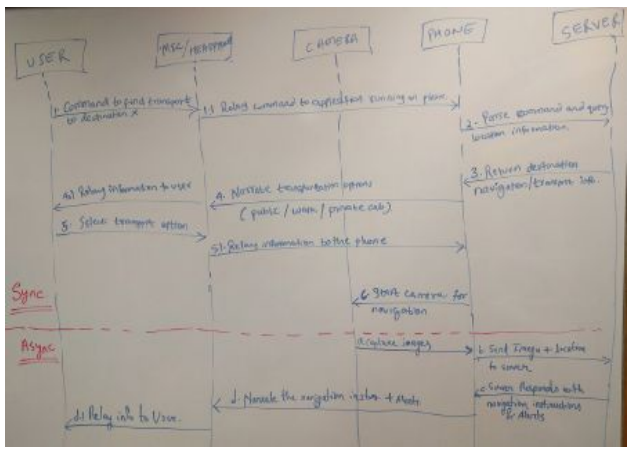

**Figure 5. Sequence diagram depicting the interaction model for Transport Assistant.**

**Use of System**
The typical use-cases that the system addresses are:

- The user will be able to order and manage his own transportation.
- User will be able to navigate to the destination or pick up point with required assistance through auditory feedback.
- There will be a Q&A feature that the user may use as a supplement to get more information about his travel or location. Eg: The user will be able to ask questions like "how is the weather outside today".
- The system will also be able to sense the close surroundings of the user. For example, a sight of zebra crossing may alert user that he is approaching a road.

**Technical Requirements**
Hardware Requirements

- Portable Camera - Bluetooth enabled or which can connect to a mobile phone via WiFi. We plan on mounting this camera on the user's body via a wearable waist belt.
- Android phone - We are currently targeting Android platform version 6.0 and above, having at least 1GB of RAM, 16GB storage, and dual-core processor.
- Headphone & Mic - Bluetooth enabled or which can connect to a mobile phone via a normal cable (3.5mm jack).

Software Requirements

- We perform image recognition based on the visual signals captured by the camera, using the Watson Visual Recognition API hosted on IBM Bluemix platform.
- To analyse the audio commands given by the user we leverage Google's speech to text service natively provided on the Android platform.
- Our server-side application is hosted on Amazon Web Services (AWS) cloud, which provides an excellent scalable and reliable cloud platform.

**PROTOTYPE**
The main goal of the prototype was to demonstrate the ability to assist a visually impaired individual through his transportation needs. First and foremost requirement was a convenient voice interactive interface using which the user would interact with the application. Apart from voice interaction following were some of the key features that we decided to include in the prototype:

- Image recognition to provide assistive vision along with assistive transportation.
- Voice augmented indoor and outdoor navigation.

Following are the scenarios that were demonstrated:

1. Hotword detection - saying "hello assistant" to trigger a reaction from the application.
2. Finding the nearest Bus Stop to the user.
3. Finding the next bus to a required destination and follow up getting more details about the bus.
4. Finding the total time required to travel to a destination by bus, car and walk.
5. Finding out about weather today and tomorrow.
6. Requesting an Uber ride to a destination.
7. Environment sensing through IBM Watson Visual Recognition.

One of the strong points of our project prototype is the eclectic mix of the technologies used. The application development has been primarily carried out on Android platform through Android Studio. We have used the native voice recognition and speech to text/ text to speech modules provided by Google and Android. To fetch the transport related navigation details, we have used Google Maps APIs whose responses are being parsed to form a meaningful sentence that is relayed back to the user through voice.

Furthermore, we deployed an instance of IBM Watson Visual Recognition Service on IBM Bluemix platform whose VIsual Recognition APIs were used to detect various objects in the surrounding. For the demonstration purpose, we specifically trained the Watson's visual recognition API to detect Zebra Crossing and generate an alert to the user that he is approaching a road crossing. We have also implemented a prototype of indoor/ outdoor navigation assistance, wherein the user is provided with walking directions and the body mounted camera would opportunistically capture images and detect obstacles and objects through visual recognition. Figure 6 shows the P2P WiFi camera which performs this task. The walking instructions sent as 'firebase' push notifications, are then relayed to the user over voice.

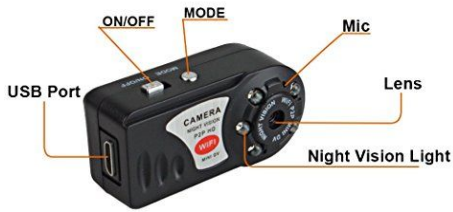

**Figure 6. FREDI Mini Portable P2P WiFi Camera**

We also hosted a sample application that mimics Uber server that receives and schedules a ride. It also sends out messages to the user and the driver. Note that the transport assistant application is programmed to read any text message received aloud. The application was hosted on an EC2 instance on Amazon AWS. Another crucial feature added was the battery low alert. A Blind person should get sufficient time to seek alternative help before the battery of the device drains off.

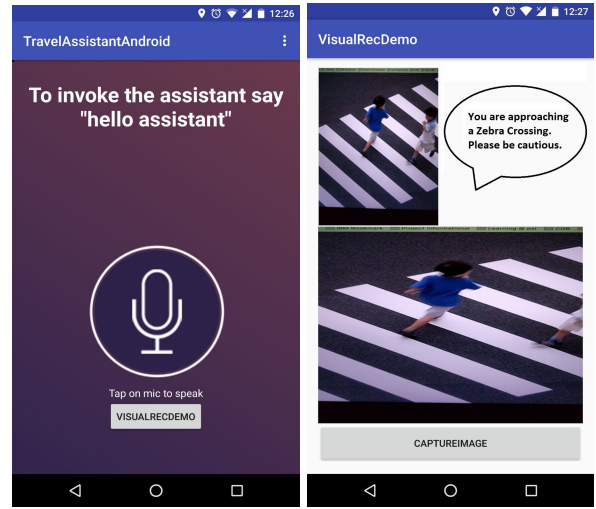

**Figure 7. Screenshots of the Transport Assistant mobile application. (Left) the home screen wherein the application listens to voice inputs and provides response. (Right) the demo for Zebra Crossing detection.**

## RESULTS

The results of Transport Assistant were pretty impressive. We were able to achieve what we envisioned; both, in terms of performance, and also in making it pervasive. The speech-to-text conversion using native Android APIs was amazingly accurate. Our application design made it possible to conveniently extend the existing set of questions which the Transport Assistant could answer, by simply inserting additional regular expressions for the grammar. The response time for fetching answers from the server hosted on AWS was relatively fast, making sure the application did not lag - another design goal we strived towards.

Below is the summary of performance and results for different functionalities supported by the Transport Assistant:

Triggering a reaction by saying "hello assistant"

- The Transport Assistant application ran silently in the background while not in use, or when the application is closed.
- When it hears the hotword "hello assistant", it promptly wakes up and responds immediately with "hello, how may I help you?".
- The reaction time is pretty fast, even when different people pronounce "hello assistant", which proves the speech to text conversion is accurate.

Getting information about public transport, weather

- Saying "hello assistant" triggers a conversation with the assistant. Subsequently the user can ask any question related to public transportation, weather, etc.
- The assistant could answer a question like, "tell me the nearest bus stop", with accurate information in minimal amount of time; in a conversational style.
- Even a complex question such as, "which is the next bus to go to Wells Library?", was answered accurately by the assistant within milliseconds by looking up *Bloomington Transit*, which is the agency that operates public transport here in Bloomington, Indiana.
- More complex questions such as "tell me details about this bus" or, "how long would it take me to get there?"; the assistant could answer with relative ease by parsing the english grammar associated with the question, and matching it against appropriate action handlers that retrieves information that is asked in a short time.

Object identification using Watson VR APIs

- The assistant was able to successfully capture images and accurately identify the object in the image by sending it to the Watson VR APIs we trained.
- We captured random images of zebra-crossings, and the assistant would correctly classify them, along with also intelligently identifying the street and giving out an alert to the user, "you are approaching a zebra crossing, please be cautious".
- We even tested with other random object images such as coffee cups, and the assistant would spot on recognize even the coffee brand. Eg: we captured an image of a Starbucks coffee cup and the assistant recognized it as "Americano Starbucks Coffee"; or a laptop as "Machine"; a human as "Person".
- The response time was pretty quick, and the assistant was able to fetch the results within matter of seconds.

Indoor navigation using Push notifications

- We used Firebase for sending out push notifications to the user, to instruct him/her to either "walk straight", "turn left", "turn right", "stop! obstacle ahead".
- The time duration between sending of the push from the server, and its reception on the phone was very small - it almost seemed instantaneous.

**CONCLUSION**

Empowering human lives have been the key goals of Pervasive technologies and through our project, we have attempted to bridge the challenges faced by visually impaired. Transportation was one area which requires due attention for the target population. We were able to successfully demonstrate how the Transport Assistant application prototype helps in solving certain key transportation problems that were realized throughout the process.

We presented our prototype at the Research Symposium for Fall 2016 conducted by the School of Informatics and Computing, Indiana University. We won the award for the "Best Graduate Research Project". The symposium was an amazing platform to showcase our research work, and to receive enormous amount of feedback from different people, especially from the people whom we interviewed who took out their valuable time and efforts to visit our booth on the day of the symposium. It is good to see the breadth of research projects and options available using technology.

We received a lot of compliments for our work, but more for the fact that we thought about leveraging today's technology for providing better services to the visually impaired. As we asserted, this is just a small step towards a bigger goal of making this world a better place to live for all people, not restricted by physical or mental limitations; thereby levelling the playing field for one and all.

**ACKNOWLEDGEMENTS**

We would like to express great gratitude to Prof. Kay Connelly and the lab instructors for all their help, support and guidance throughout the project. This work would not have been possible without their help. We would also like to thank the School of Informatics and Computing for providing us with the infrastructure and hardware to work on this project, and for allowing us to participate in the Fall Research Symposium. In addition, we would like to sincerely thank Tousif Ahmed (Eshan) for his invaluable guidance and expert opinions. Shirley Stumpner for providing informative and encouraging pointers. Finally, a big thank you to all those who volunteered to interview with us and sharing their problems and thoughts.